\documentclass[12pt]{article}
\usepackage{graphicx}
\usepackage{amssymb}
\graphicspath{{./}{./}}
\usepackage[authoryear]{natbib}
\newcommand{\captionfonts}{\normalsize}

\makeatletter  
\long\def\@makecaption#1#2{%
  \vskip\abovecaptionskip
  \sbox\@tempboxa{{\captionfonts #1: #2}}%
  \ifdim \wd\@tempboxa >\hsize
    {\captionfonts #1: #2\par}
  \else
    \hbox to\hsize{\hfil\box\@tempboxa\hfil}%
  \fi
  \vskip\belowcaptionskip}
\makeatother   

\begin{document}

\hspace{13.9cm}

\ \vspace{20mm}\\

{\Large Intrinsic adaptation in autonomous recurrent neural \\networks}

\ \\
{\bf \normalsize Dimitrije Markovi\'c$^{1}$ and Claudius Gros$^{1}$}\\
{$^{1}$Institute for Theoretical Physics, J. W. Goethe University, Frankfurt am Main, Germany.}\\
%


\thispagestyle{empty}
\markboth{}{NC instructions}
\ \vspace{-0mm}\\


\begin{abstract}
{\normalsize A massively recurrent neural network responds on one 
side to input stimuli and is autonomously active, on 
the other side, in the absence of sensory inputs. 
Stimuli and information processing depends crucially 
on the qualia of the autonomous-state 
dynamics of the ongoing neural activity. This default
neural activity may be dynamically structured in time 
and space, showing regular, synchronized, bursting 
or chaotic activity patterns.

We study the influence of non-synaptic plasticity 
on the default dynamical state of recurrent neural 
networks. The non-synaptic adaption considered acts 
on intrinsic neural parameters, such as the threshold and 
the gain, and is driven by the optimization of the 
information entropy. We observe, in the presence of 
the intrinsic adaptation processes, three distinct and 
globally attracting dynamical regimes, a regular synchronized, 
an overall chaotic and an intermittent bursting regime. The 
intermittent bursting regime is characterized by intervals of 
regular flows, which are quite insensitive to external stimuli, 
interseeded by chaotic bursts which respond sensitively to 
input signals. We discuss these finding in the context of 
self-organized information processing and critical brain dynamics.}
\end{abstract}

\section{Introduction}

In the last couple of decades self organized processes have 
attracted the interests of many researchers from various 
scientific areas, both in natural and in social sciences. 
A system is said to be self-organizing, quite generally, 
when a state of high dynamical complexity arises
reliably from relatively simple basic organization rules
\citep*{Ashby1962,Camazine2003,Gros2010}. 

It is often the case that the self-organization in dynamical 
systems is achieved through an interplay or regulative forces
involving positive and negative feedback, {\it viz.} through the
interplay of internal drives which act destabilizing and 
regulating, respectively, onto the dynamics of the system. 
In general one type of feedback can dominate, driving the
system towards a chaotic or towards an ordered phase respectively. 
A proper balance of the two opposing drives can bring 
the dynamical state at the point of a phase transition, a
critical state. One speaks of self-organized criticality (SOC) 
whenever this balance is not achieved through the actions of an 
outside controller but through internal self-organizing processes
\citep*{Bak1987,Bak1988,Bak1995,Adami1995}.

As a dynamical system approaches a critical point, its 
spatiotemporal complexity rises. It has been suggested
that this rise in the complexity  improves the computational 
properties and the capability of dynamical systems to process
information \citep*[e.g.,][]{Sole1995,Bertschinger2004a,Legenstein2006}. 
This notion of {\em ``Computation at the edge of chaos''} may
also been seen in the broader context of
{\em ``Life at the edge of chaos''} \citep*[see][]{zimmer99,Gros2010};
the dynamical systems at the underpinning of all living
have the tendency to self-organize themselves close to
a critical state.

In recent years there have been many studies of 
the possible occurrence of SOC in neural networks 
with synaptic plasticity. Most of these studies
have concluded that synaptic plasticity drives the 
dynamics generically far below a critical point 
\citep*[e.g.,][]{Siri2007a,Siri2008,Dauce1998}, {\it viz.} it over-regulates 
the network dynamics. Hence, an organizational principle 
is needed which will maintain an intermediate level of excitability 
in neural networks, preventing the occurrence of dynamical states 
which are non- or hyper-reactive to external influences.

It has been assumed for many years that the dominant driving force, 
shaping the brain's dynamical state, is the synaptic plasticity.
Thus, little attention was put to other forms of neural adaption,
the non-synaptic adaptation of individual neurons 
\citep*{Mozzachiodi2010}, also known as intrinsic plasticity. 
Intrinsic plasticity is mostly manifested as a change in the 
excitability of a neuron, where this change is achieved through 
the adaptation on the level of membrane components. Here, we 
investigate the role of non-synaptic
plasticity in the formation of complex patterns of 
neural activity. 

We study a previously proposed model of intrinsic plasticity 
\citep*[see][]{Triesch2005,Triesch2007,Stemmler1999},
and its influence on the dynamical properties of autonomous
recurrent neural networks with rate encoding neurons 
in discrete time. Within this model neurons aspire to achieve, as
an average over time, a firing-rate distribution which
maximizes the Shannon information entropy \citep{Gros2010}. 
Therefore, the neurons are trying to homeostatically regulate 
an entire distribution function, a mechanism denoted 
polyhomeostatic optimization \citep*{Markovic2010}.

Intrinsic plasticity in the form of polyhomeostatic optimization
gives rise, for random recurrent network topologies, to ongoing and
self-sustained neural activities with non-trivial dynamical states.
Depending on the target mean firing rate and network parameters, 
one observes three distinct phase states: synchronized oscillations, 
an intermittent-bursting and a chaotic phase; all states being 
globally attracting in their respective phase spaces.

In Section \ref{sec_adaptation} we derived 
the stochastic learning rules for intrinsic 
adaptation. This is followed by the analysis of
a single self-coupled neuron with intrinsic 
plasticity (Section \ref{sec_neuron}) 
and the analysis of recurrent network with intrinsic 
plasticity (Section \ref{sec_network}). Concluding remarks 
and discussion are finally provided in Section \ref{sec_discussion}.

\section{Stochastic adaptation}\label{sec_adaptation}

We used a basic discrete-time, rate-encoding artificial 
neuron model. The firing rate $y\in[0,1]$ of the neuron is given 
as a nonlinear transformation of the total synaptic 
input current $x\in (-\infty, +\infty)$, $y=g(x)$. The 
transfer function $g$ has a sigmoidal form,
a usual choice being the logistic function
\begin{equation}
\label{eq_logistic}
 y(t+1) = g_{a,b}(x(t)),\qquad g_{a,b}(z) = {1 \over 1+e^{-az-b}}~,
\end{equation}
where $a$ is the gain and $b$ the bias. The parameters 
of the transfer function, intrinsic parameters, will 
eventually become slow variables with a stochastic 
learning rules determining their time evolution. 

Let us denote with $p_x(x)$ the probability density function (PDF) 
of the total input. Given the relation (\ref{eq_logistic}) between the 
input current $x$ and the output activity $y$ we find 
\begin{equation}\label{pdf_relation}
p_{a,b}(y) = \int_{-\infty}^{+\infty}\delta(y-g_{a,b}(x))\,p_x(x)\,\textrm{d}x 
= {p_x(x) \over g^\prime_{a,b}(x)}|_{x=g^{-1}_{a,b}(y)}
\end{equation}
for $p_{a,b}(y)$, the PDF of the firing rate.
The main idea behind the derivation of 
adaption rules for the intrinsic parameters $a$ and
$b$ is the assumption that the neuron's excitability 
should change in a way which maximizes the entropy 
of the firing rate distribution $p_{a,b}(y)$,
keeping at the same time the average activity 
at a desired level \citep{Triesch2005}.

The rational for this procedure is the following: 
the maximization of the firing rate entropy
implies that a neuron will use the entire range of 
available activity states, optimizing the
information transfer between neural input and output.
Furthermore, the regulation of the average firing 
rate is present due to environmental constraints 
on the neuron, \textit{e.g.} the limited 
energy resources needed for metabolic processes. 

Having a positive-definite variable $y$, with a fixed 
first moment, the maximum entropy 
PDF corresponds to the exponential distribution
\begin{equation}
 \label{max_entropy}
 p_{\lambda}(y) = \frac{1}{Z(\lambda)}e^{-\lambda y},
\qquad\quad y\in[0,1]~,
\end{equation}
where $Z(\lambda) = \int_0^1 e^{-\lambda y}\textrm{d}y$ is 
the partition function. We will refer to the first moment 
of the PDF (\ref{max_entropy}), denoted as $\mu$, as the 
target average firing rate. It is given by
\begin{equation}
 \label{max_mean}
\mu = \int_0^1 yp_{\lambda}(y)\textrm{d}y = 
\frac{1}{\lambda}-\frac{1}{e^{\lambda}-1}~.
\end{equation}
In general the inverse function $\lambda(\mu)$ cannot
be found, but for $\lambda\gg1$ we recover the 
$\lambda\to1/\mu$, which is valid for
exponential PDFs defined on $[0,\infty]$.

A natural way to introduce a distance measure between
two PDFs is the Kullback-Leibler (KL) divergence
\citep{Gros2010}, defined as
\begin{eqnarray}
 \label{eq_kl_divergence}
\nonumber
D_{\lambda}(a,b) &=& \int p_{a,b}(y)
\ln\left(\frac{p_{a,b}(y)}{p_{\lambda}(y)}\right)\textrm{d}y\\
&=& - E_{p_x}[\ln g^\prime_{a,b}(x)] + \lambda E_{p_x}[g_{a,b}(x)] -H[p_x] + \ln Z(\lambda)~,
\end{eqnarray}
where $E_{p_x}[\cdot]$ denotes the expectation value 
with respect to $p_x(x)$, and $H[p_x]$ denotes 
the differential entropy functional. By minimizing
$D_{\lambda}(a,b)$ with respect to the intrinsic parameters
$a$ and $b$ one obtains the learning rules. In 
Eq.~(\ref{eq_kl_divergence}) only the first two terms 
are functions of $a$ and $b$. The gradient descent 
hence gives the following relation
\begin{eqnarray}
\nonumber
-\frac{\partial D_\lambda(a,b)}{\partial\alpha}
&=& E_{p_x}\left[{\partial \over \partial\alpha}
\ln g^\prime_{a,b}(x) - \lambda{\partial \over \partial\alpha}g_{a,b}(x)\right] \\
& =& \int p_x(x)
\underbrace{\left[{\partial \over \partial\alpha}
\ln g^\prime_{a,b}(x) - \lambda{\partial \over \partial\alpha}g_{a,b}(x)\right]
}_{\equiv\, \Delta_\alpha}\textrm{d}x~,
 \label{grad_disc}
\end{eqnarray}
with $\alpha \in \{a, b\}$. As the input distribution 
$p_x(x)$ is in general unknown, it is suitable 
to derive the adaption rules by using a stochastic 
gradient descent \citep{Spall2005}. Such adaption rules, 
for the update of the internal parameters $a$ and $b$, 
are obtained by using the expression between the brackets
on the right-hand side of Eq.~(\ref{grad_disc}) .
An advantageous side effect of this approach is that 
the adaptation rules become local 
in time. Using Eq.~(\ref{eq_logistic}) for
the transfer function $g_{a,b}(x)$ to evaluate
$\Delta_a$ and $\Delta_b$, we obtain the stochastic 
learning rules

\begin{eqnarray}
\nonumber
a(t+1) &=& a(t) +\epsilon\Delta_a(t)
\ =\ a(t) + \epsilon\,\big(1/a(t) + x(t)\Delta(t)\big)\\
b(t+1) &=& b(t) +\epsilon\Delta_b(t)
\ =\ b(t) + \epsilon\, \Delta(t)~,
\label{eq_learning}
\end{eqnarray}
where $\epsilon$ is the learning rate and 
$$
\Delta(t) = 1 - (2+\lambda)y(t+1)+\lambda y(t+1)^2~.
$$ 
The learning rate $\epsilon$ is assumed to 
be small; viz the time evolution of the internal parameters 
is slow compared to the evolution of both $x(t)$ and $y(t)$. 
In this way the stochastic adaptation, which depends only on 
the instantaneous values of the variables, can closely match the 
direction of the deterministic gradient Eq.~(\ref{grad_disc}). 
Also, the input distribution can, in general, be non-stationary, 
therefore the minimum of the cost function $D_{\lambda}(a,b)$ 
could vary in time. For this reason, the learning rate should 
also be large enough for the adaptation to follow the changing minimum.  
However, any finite and constant learning rate $\epsilon>0$
doesn't satisfy the condition for exact convergence of internal 
parameters into the minimum of Kullback-Leibler divergence 
\citep{Spall2005}. Still, if the learning rate decreases 
with every time step, a condition needed for strict convergence, 
the intrinsic adaptation will react slower with time to a 
variability in the position of the minimum of the 
cost function. Thus, it's more favorable to have here a constant 
learning rate which will result in the oscillations of the 
intrinsic parameters around the minimum. The amplitude of this 
oscillations roughly scales as $\epsilon$ \citep{Bottou2004}, thus a 
small learning rate also ensures convergence within a small vicinity 
from the minimum of the cost function $D_{\lambda}(a,b)$.

\begin{figure}[t]
\centerline{
\includegraphics[width=\textwidth]{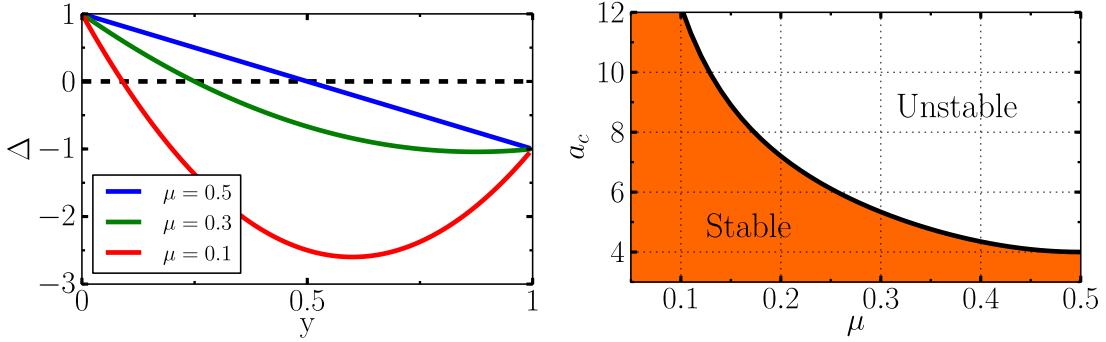}
}
\caption{(Left) A dependence of relative parameter change 
$\Delta$ (Eq.~\ref{eq_Delta_tile_b}) on output activity 
$y$ for different target firing rates $\mu$. 
(Right) Critical value of the gain $a_c$, as 
a function of the average firing rate $\mu$. The colored area 
shows the region of stability of the fixpoint 
$(y^*(\lambda), b^*(a,\lambda))$, see Section \ref{sec_neuron}.
}
\label{fig_stability}
\end{figure}

\section{Single neuron}\label{sec_neuron}
We analyze initially a minimal network setup, 
a self-coupled neuron adapting homeostatically the
intrinsic parameters of the transfer function. 
A synaptic connection between the axon and the dendrites 
of the same neuron is also known as an autapse.

Neurons with an autapse are not rare in the brain. 
They have been observed in various brain regions 
and in different types of neurons. The discovery 
of functional autapses provides clues for possible 
physiological roles \citep{Bekkers2003}. 
\citet*{Herrmann2004} suggest that autapses lead 
to oscillatory behavior in otherwise non-oscillating 
neurons. We shall see below how this type of behavior 
spontaneously arises in self-excitatory neurons with 
intrinsic plasticity. We focused on the analysis of 
excitatory autapse and used it as a basis for understanding 
the observed behavior in a larger network setup 
(see Section \ref{sec_network}). For the case of 
self-inhibition please refer to a separate study \citep{Markovic2010}.

The autapse neuron is equivalent to the identification
$x\to y$ in Eqs.~(\ref{eq_logistic})
and (\ref{eq_learning}). The complete set of evolution
rules for the dynamical variables $y(t),\, a(t)$ and $b(t)$ is then
\begin{eqnarray}\label{eq_one_site}
\nonumber
y(t+1) & =& g_{a(t),b(t)}\big(y(t)\big) \\
b(t+1) & = & b(t)+\epsilon\Delta(t) \\
a(t+1) &=& a(t)+\epsilon (1/a(t)
             + y(t)\Delta(t))
\nonumber
\end{eqnarray}
with
\begin{equation}
\Delta(t)\ =\ 1-(2+\lambda)y(t+1)+\lambda y^2(t+1)~.
\label{eq_Delta_tile_b}
\end{equation}
The right-hand side of (\ref{eq_Delta_tile_b})
depends directly on $y(t+1)$ and only implicitly on $y(t)$, as
one can easily verify when going through the derivation
of the rules (\ref{eq_learning}) for the intrinsic plasticity.
In plot at the left side of Fig.~\ref{fig_stability} we showed
$\Delta(y)$ (Eq.~(\ref{eq_Delta_tile_b})) for various target 
firing rates $\mu$. Note that $\Delta(y=0)=1$ and
$\Delta(y=1)=-1$, independently of $\lambda$.
%

\begin{figure}[t]
\centerline{
\includegraphics[width=0.8\textwidth]{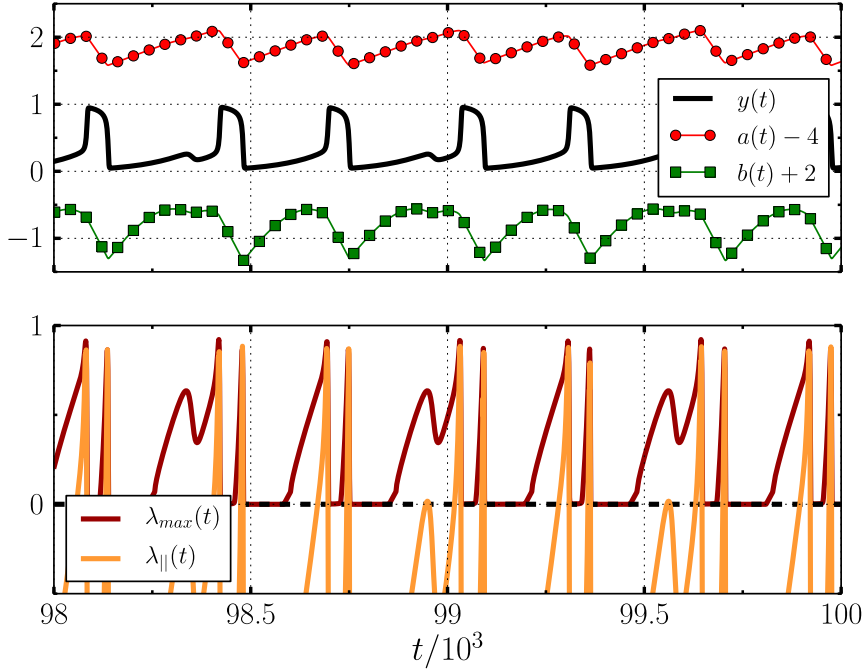}
           }
\caption{(Top) The time dependence of firing-rate $y(t)$ (solid line), 
bias $b(t)$ (squares) and gain $a(t)$ (circles) for
the one-site problem (\ref{eq_one_site}), with a learning 
rate $\epsilon=0.01$ and a target average firing rate $\mu=0.25$;  
The gain $a(t)$ is set initially bellow a critical value  
and since $\epsilon\ll1$ the system relaxes quickly to the fixpoint
of $y=g_{a,b}(y)$. Once $a(t)$ surpasses a certain threshold,
compare Fig.\ \ref{fig_stability}, the fixpoint becomes unstable 
and the system follows a limiting cycle.\newline 
(Bottom) Maximal local Lyapunov $\lambda_{max}(t)$ exponent 
compared to a Lyapunov exponent of a perturbation 
parallel to the flow $\lambda_{||}(t)$. They were estimated along the 
points of the trajectory ($y(t),a(t),b(t)$).
}
\label{fig_oneSiteTime}
\end{figure}

\subsection{Stability analysis}
We first analyze a reduced model of the three 
evolution equations (\ref{eq_one_site}), obtained by 
setting $\Delta a(t) = 0$, viz considering
a constant $a=a(t)=a(t+1)$. The reduced system contains 
a fixpoint $(y^*(\lambda), b^*(\lambda,a))$, where 
$y^* = [(2+\lambda)-\sqrt{4+\lambda^2}]/2\lambda$ and 
$b^* = \ln(y^*/(1-y^*))-ay^*$. This fixpoint defines a one 
dimensional manifold in the complete phase space 
$(y,b,a)$, where the stability of the manifold depends 
directly on the given value of $a$ and $\lambda$ 
(see Fig.~\ref{fig_stability}). For $a<a_c$ the 
dynamics is attracted toward the fixpoint $(y^*, b^*)$,
while for $a>a_c$, the fixpoint becomes repelling and 
the activity of the neuron follows a limiting cycle. 
One can show that the critical gain is
$a_c = 1/y^*(1-y^*)$.

The time evolution of the full set of equations 
(see Fig.~\ref{fig_oneSiteTime}, top) approaches a 
limiting cycle, for all starting values of $(y,b,a)$.
The evolution rules (\ref{eq_one_site}) have
fixpoint solutions also for a vanishing adaptation,
\textit{viz.} for $\epsilon\to0$. These fixpoints 
are turned for $\epsilon>0$ into attractor relics 
\citep{Gros2007a,Gros2009}. The trajectory slows down 
close to the attractor relics, giving rise to the 
transient firing states, observable in Fig.~\ref{fig_oneSiteTime}. 
This non-trivial activity pattern is a direct consequence
of the polyhomeostatic adaption principle. The system
cannot achieve, as an average over time, a non-trivial
firing-rate distribution by settling into a steady state.
Polyhomeostatic adaption hence forces the neuron to remain
autonomously active, with varying firing rates.
%

\begin{figure}[t]
\begin{center}
\includegraphics[width=0.8\textwidth]{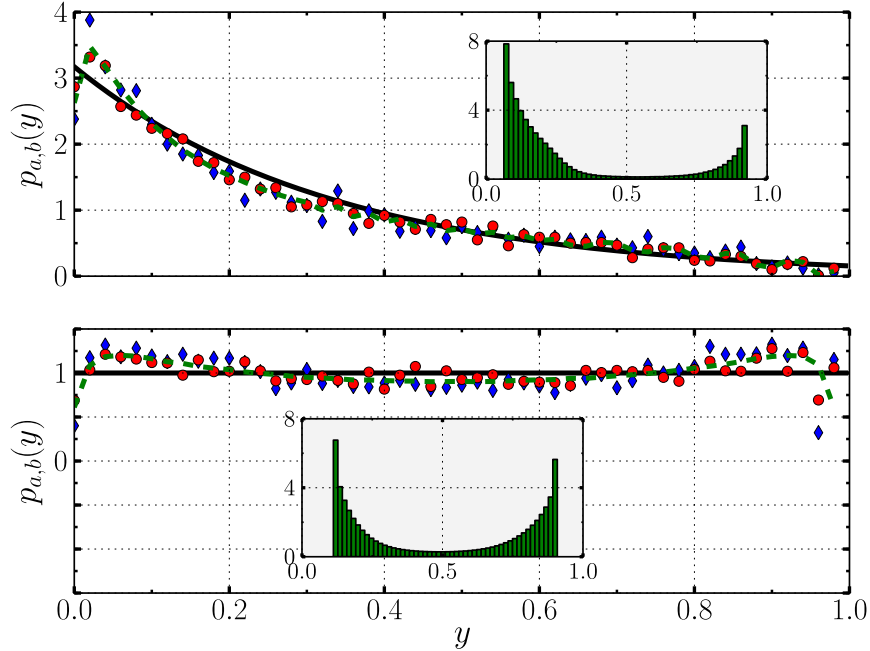}
\end{center}
\caption{Output distributions of the two neurons
with highest (diamonds) and lowest (circles) 
Kullback-Leibler divergence (\ref{eq_kl_divergence})
compared to the mean output distribution 
(dashed  line) and the target exponential 
output distribution (full line). The network size 
$N=500$ neurons and a target mean firing 
rate $\mu = 0.28$ (top) and $\mu = 0.5$ (bottom). 
The values are identical for all the neurons in the network
and fraction of excitatory links $f_{exc}=0.5$. 
Insets: Output distribution of the single self-coupled 
neuron having the target average firing rate $\mu$ at the 
same value as the neurons in the network.
}
\label{fig_distr} 
\end{figure}

For an insight on the influence of external input signal
on the dynamics, we have estimated the maximal local Lyapunov 
exponent $\lambda_{max}(t)$ and the Lyapunov exponent for a 
perturbation in the direction of the flow $\lambda_{||}$ ($\delta\vec{r}(t) 
\propto (\Delta y(t),\Delta a(t), \Delta b(t))$). They are 
presented in the bottom graph of Fig.~\ref{fig_oneSiteTime}. 
We see that the neuron is most sensitive to a perturbation 
during the transition between two attractor relics (low and 
high activity levels), since $\lambda_{max}$ is positive through 
these transition periods. Also, $\lambda_{||}\approx \lambda_{max}$
during the fast transition between the attractor relics. We
thus conclude that the direction of maximal sensitivity to perturbations 
is aligned with the direction of the flow at this points.  
Note that the two attractor relics can be stable 
during the same time period, although the activity settles 
in only one of them. This means that a transition could be
induced with a sufficiently strong perturbation. 
%

\begin{figure}[t]
\centerline{
\includegraphics[width=1\textwidth]{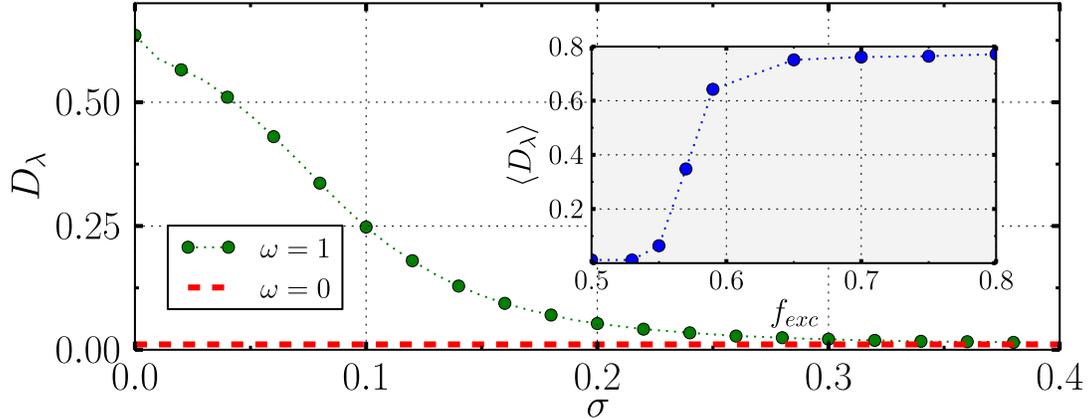}
           }
\caption{The Kullback-Leibler (KL) divergence $D_\lambda(a,b)$
of the neuronal firing-rate distribution relative to
the target exponential distribution (\ref{max_entropy}),
as a function of the standard deviation $\sigma$
of a Gaussian input distribution $p_x(x)$,
in the presence of an autapse (green dots, $w=1$
in Eq.~(\ref{autapse_external})) and in the absence
of the autapse (red dashed line, $w=0$ in Eq.~(\ref{autapse_external})).
Target mean firing rate $\mu$ is set to $0.3$ .
Inset: Mean KL divergence $\langle D_{\lambda} \rangle$ 
(see section \ref{sec_network}) 
of the random recurrent neural network with $N=1000$ neurons and mean 
target firing rate $\mu=0.3$, as a function of the fraction of 
excitatory links $f_{exc}$. Note that increase of $f_{exc}$ can be 
related to the decrease in the noisy component of the input that 
each neuron receives (see section \ref{sec_oscibih}). 
}
\label{fig_klOutputNoise}
\end{figure}

\subsection{Noisy autapse}
Let us consider the case of a noisy autapse, when 
\begin{equation}
\label{autapse_external}
x(t) \rightarrow w y(t)+\xi(t)~.
\end{equation}
The neuron receives, beside the autaptic signal $\sim wy(t)$, 
a random input from an external source $\sim \xi(t)$ (e.g.\ 
from some other neurons in the network). The non-autaptic component 
of the input is drawn from a Gaussian distribution, 
$\xi\sim\textit{N}(0,\sigma)$, where $\textit{N}(0,\sigma)$ 
denotes a normal distribution with zero mean and 
variance $\sigma^2$.

The external input hence perturbs the signal coming from 
the autapse. From Fig.~\ref{fig_oneSiteTime} it is quite 
obvious that the output distribution of a self-coupled
neuron with $\xi\equiv0$, {\it viz.} noiseless autapse, 
in Eq.~(\ref{autapse_external}), deviates substantially
from the target exponential distribution. 
The output distribution in the case of noiseless 
autapse is presented in the insets of Fig.~\ref{fig_distr}.  
However, as we increase the magnitude of the external 
signal, the output distribution of the neuron
approaches the optimal distribution and the KL
divergence decreases toward a minimum, see
Fig.~\ref{fig_klOutputNoise}. Obviously, when 
$\sigma \gg \omega$, the external input dominates, and the 
two cases with and without an autapse become equivalent. 
Nevertheless, even small amounts of noise, that is $\sigma < \omega$ are 
sufficient to disrupt the oscillatory behavior of the output 
activity. This happens because of the existence of 
the second stable fixpoint, for certain values of 
$a$ and $b$ in $y = g_{a,b}(y)$. As the standard deviation
$\sigma$ increases, the probability that the firing-rate $y$ will 
transit toward the second fixpoint also increases. 
Thus, at a certain levels of noise, the activity stochastically 
escapes in short time intervals from the stable fixpoints, and 
the regular oscillatory behavior is destroyed.
This also implies that a certain level of decorrelation 
between the input current and output activity has 
to be reached, if the firing-rate distribution is 
to come as close as possible to the desired target 
distribution. 
%

\begin{figure}[t]
\centerline{
\includegraphics[width=1\textwidth]{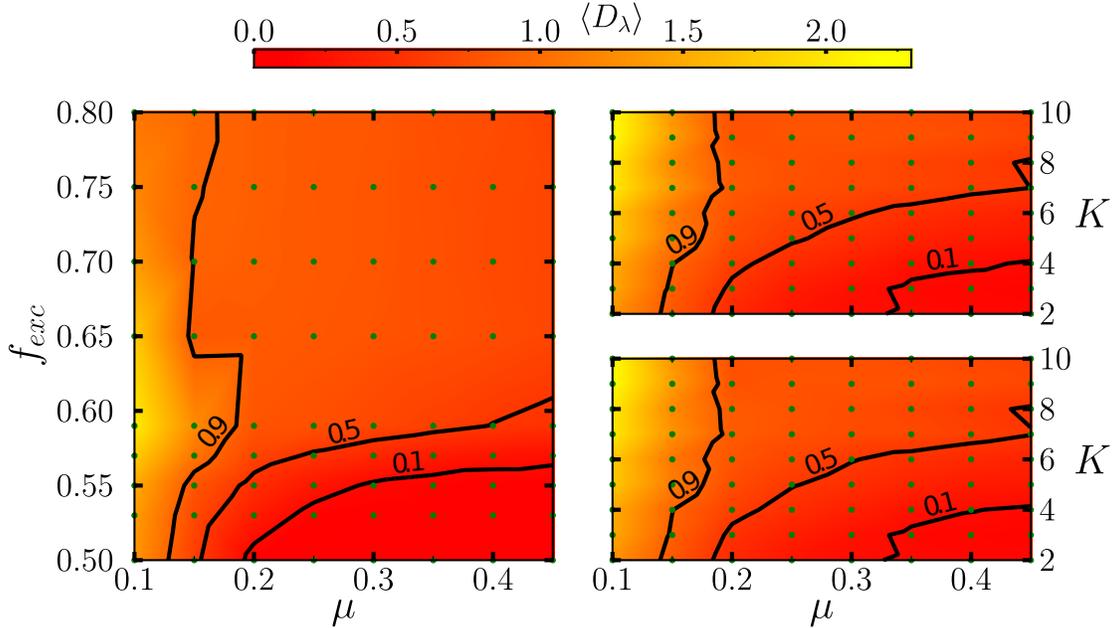}
}
\caption{(Left) Mean Kullback-Leibler (KL) divergence 
$\langle D_{\lambda}\rangle$ (color coded)
as a function of the fraction of excitatory links $f_{exc}$ and
target mean firing rate $\mu$ in a recurrent network of $N=1000$ 
neurons and connectivity $K=100$. (Right) $\langle D_{\lambda}\rangle$ 
as a function of connectivity $K$ and target mean firing rate $\mu$, 
with excitatory/inhibitory neurons (Top) and projections (Bottom). 
Fraction of excitatory links $f_{exc} = 0.8$ in both cases.
Density plot was evaluated as a linear interpolation of the experimentally 
obtained values represented with green dots.  
}
\label{fig_meanKLlinks}
\end{figure}

\section{Recurrent neural network}\label{sec_network}

We studied numerically random recurrent neural
networks (RRNN) of $N$ polyhomeostatically 
adapting neurons (\ref{eq_learning}), where each 
neuron receives input from $K$ pre-synaptic neurons. 

In a first step we consider networks of dual neurons, 
i.e. a single neuron can have both excitatory 
and inhibitory projections. In such setup, 
the synaptic input that the $i$th neuron receives 
is expressed as
\begin{equation}
x_i(t) \ =\ {\displaystyle\sum\limits^K_{j\ne i}} w_{ij} y_j(t)~.
\label{eq_couplings}
\end{equation}

The synaptic weights are selected as $w_{ij}=\pm 1/\sqrt{K}$,
with a probability $f_{exc}$ for the weight to be positive.
The learning rate in (\ref{eq_learning})
is set to $\epsilon=0.01$. We consider
homogeneous networks where all neurons have identical 
target average firing rate $\mu$, determined through (\ref{max_mean}),
for the target output distributions 
$p_\lambda(y)$, see Eq.~(\ref{max_entropy}).

For a neuron to ideally map an arbitrary input to the 
exponential distribution with specified mean, it needs 
a transfer function which can take any functional form 
during the adaption process. This is obviously not the case 
for the logistic function which has only two adaptable parameters. 
As a result of this limited flexibility of the transfer function, 
even when the input of the neuron is independent from the 
output (Fig.~\ref{fig_klOutputNoise} red dashed line), 
the output distribution will never ideally match the 
target exponential distribution \citep{Markovic2010,Triesch2005}. 

\subsection{Dynamical behaviors}
To examine the mean deviation of the output activity 
from the target exponential distribution, we have estimated the
KL divergence $D_{\lambda}(a_i,b_i)$, see Eq.~(\ref{eq_kl_divergence}), 
for all neurons in a network of $N=1000$ neurons, and 
averaged over the entire network and over $n=50$ random 
network realizations. The obtained mean $\langle D_{\lambda} \rangle$ 
is presented in Fig.~\ref{fig_meanKLlinks} 
as a function of target mean firing rate $\mu$, fraction
of excitatory links $f_{exc}$ and network connectivity $K$.
Note that  $\langle D_{\lambda} \rangle$ is low for high 
target firing-rates and for balanced excitation/inhibition, 
or for low connectivity $K$. 
%

\begin{figure}[t]
\centerline{
\includegraphics[width=1\textwidth]{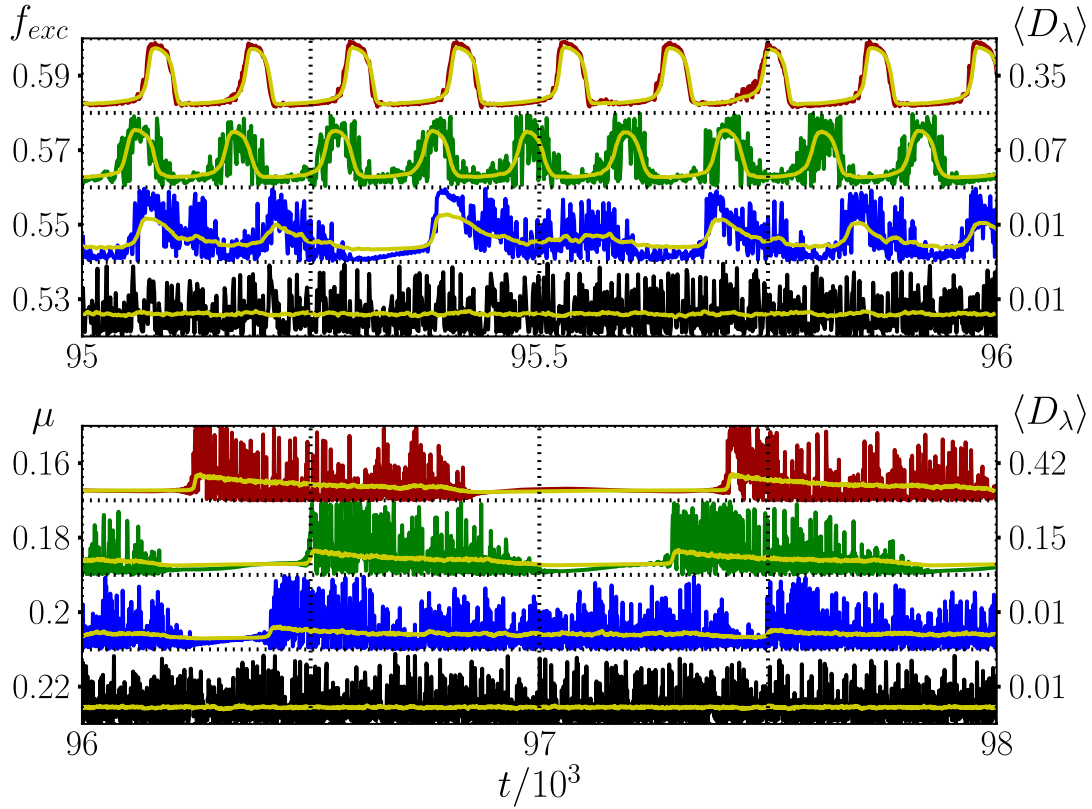}
}
\caption{Activity of one, randomly chosen, neuron 
from the network of N=1000 neurons with connectivity K=100, 
depending on the fraction of excitatory links $f_{exc}$ (top; $\mu = 0.3$) 
and the mean target firing rate $\mu$ (bottom; $f_{exc}=0.5$), where the 
yellow line represents average network activity. The right ordinate 
shows the corresponding mean Kullback-Leibler divergence 
(see section \ref{sec_network}).}
\label{fig_acti}
\end{figure}

We have observed three distinct dynamical regimes, 
a pure chaotic regime characterized by low values 
of $\langle D_{\lambda} \rangle$, a synchronised 
oscillatory regime observed in random networks with 
dominating excitatory connections and a 
intermittent-bursting regime observed for balanced 
excitation/inhibition and small $\mu$.  

To illustrate the difference between these dynamical 
behaviors, we present in Fig.~\ref{fig_acti} the average 
neural activity (yellow line) and the activity 
patterns of a randomly selected neuron in the 
network of $N=1000$ units. As we vary the fraction 
of excitatory links $f_{exc}$ and keep $\mu$ fixed 
(Fig.~\ref{fig_acti}~Top) the dynamics shifts from 
the chaotic phase into the phase of synchronised 
oscillations. On the other hand, reducing the target 
mean firing rate for balanced excitation/inhibition 
(Fig.~\ref{fig_acti}~Bottom) leads to a manifestation 
of bursts of chaotic activity alternated by periods 
of nearly constant activity. On the right side of the 
graphs we give the values of the corresponding mean 
Kullback-Leibler divergence $\langle D_{\lambda}\rangle$.
In addition, from the oscillatory regime it is possible to 
transit back into a chaotic or intermittent-bursting regime 
(depending on the value of $\mu$) by reducing the network 
connectivity $K$. This can be easily seen from the 
similarity of the density plots on the right and left 
hand side in Fig.~\ref{fig_meanKLlinks}. 

In Fig.~\ref{fig_distr} we give an example of the 
output distributions for two neurons, with highest 
and lowest values of $D_\lambda$, when the dynamics of 
the neural network is set in chaotic regime. The two 
output distribution are compared with a corresponding 
target exponential distribution.

Alternatively to randomly selecting a single link as
excitatory or inhibitory, one can consider a case
when a single neuron is selected as either excitatory
or inhibitory. Thus, all the projections from one neuron
are of the same type. We have shown $\langle D_{\lambda}\rangle$ 
for such case on the upper right graph of Fig.~{\ref{fig_meanKLlinks}}. 
The absence of a visible difference between the upper and the 
lower graph indicates that the intrinsic adaptation, in the 
low $K$ limit, leads to the same dynamical behavior indipendent 
from having excitatory and inhibitory neurons separeted or 
neurons with both types of projections.

\subsection{Oscillatory behavior}\label{sec_oscibih}
The network dynamics makes a transition into a synchronized oscillatory 
regime (see Fig.~\ref{fig_acti}), as $f_{exc}$, the fraction  
of excitatory links, is increased. To better understand this 
oscillatory behavior let us recall the discussion from the previous 
section. In the case of a single self-coupled neuron we showed how 
a certain level of decorrelation, between the output activity and the 
input signal has to be achieved in order for a neuron activity 
to properly match the target distribution (see Fig.~\ref{fig_klOutputNoise}). 
The same argument holds in the case of RRNN. Thus, when the input is 
uncorrelated with its output activity (corresponding to the 
$f_{exc}\gtrsim 0.5$), the output distribution $p_{a,b}(y)$ 
closely matches the target distribution $p_{\lambda}(y)$. 

The total synaptic input a neuron receives can be divided
into two components. The first component is correlated
with its own output activity via excitatory recurrent
connections. The second component corresponds to the noisy
and uncorrelated part of the input which results from the
competition between inhibition and excitation.
The first correlated part of the input becomes
dominant over the noisy second contribution
as we increase the fraction of excitatory links $f_{exc}$.
The activity therefore starts to follow an oscillatory
locked-in trajectory for large fractions $f_{exc}$.

In the inset of Fig.~\ref{fig_klOutputNoise} we present
the change of mean KL divergence as the number 
of excitatory connections grows, but for a fixed 
target mean firing rate. We can see that $\langle D_{\lambda} \rangle$ 
increases rapidly once the number of excitatory 
connections starts to dominate. Note that the transition 
between two phases occurs in a slightly different manner 
compared to the case of the neuron with a noisy 
autapse. One reason for this difference is that 
input/output correlations will be amplified by 
additional delayed components of an excitatory 
feedback a neuron receives. This reasoning 
can be shown to hold by simulating a single neuron 
with delayed coupling autapses driven by the input
$x(t) \to \sum_{k=0}^{n-1} \omega_ky(t-n) + \xi(t)$~.  

\subsection{Intermittent bursts of chaotic dynamics}

In the second graph (Fig.~\ref{fig_acti}~Bottom), 
the dynamics enters an intermediate phase, characterized
by intermittent bursting of chaotic neural activities,
as the target mean firing rate is decreased. 
A closer look into the phase space of intrinsic parameters 
$(a_i,b_i)$, of the $i$th neuron, reveals that the
intrinsic parameters approach a limiting cycle, 
similar to the case of a neuron with an autapse. 
During the regime of nearly constant activity 
$\Delta_i(t) \approx 0$ the gain steadily increases.
Once the gains of sufficient number of neurons crosses
a certain critical value, the activity of the entire 
network shifts into a chaotic regime. The activity 
during the chaotic regime exceeds the target 
average activity level $\mu$, thus the 
gains of all neuron are driven back to sub-critical 
values. Even when reducing the learning rates by several 
orders of magnitude this intermittent-bursting 
behavior persists. Nevertheless, when considering
constant and supercritical gains for all neurons 
and allowing only the respective biases to adapt, 
we observe a pure chaotic behavior.
This change, which arises when reducing the
number of degrees of freedoms by considering constant
gains $a_i$, is not yet fully understood. A one possible cause
could be the use of ``vanilla'' gradient which doesn't 
take into account the curvature of the manifold 
of probability distributions $p_{a,b}(y)$ \citep[see][]{Amari}, 
and therefore doesn't point into direction of maximal change 
of $D_\lambda(a,b)$.  

\subsection{Sensitivity to external perturbations}\label{sec_ftle}

\begin{figure}[t]
\centerline{
\includegraphics[width=1\textwidth]{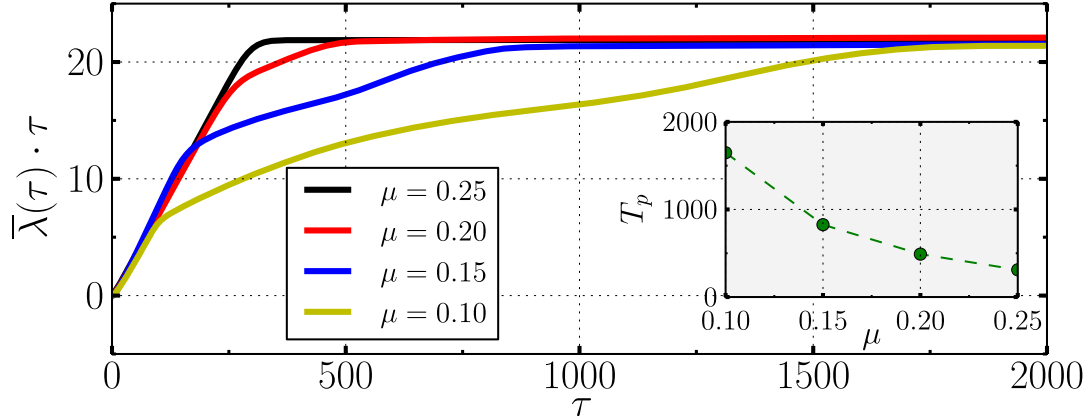}
}
\caption{Nonlinear finite time Lyapunov exponent $\bar{\lambda}(\delta_0,\tau)$ 
(see section \ref{sec_ftle}) at the time step $\tau$, for the various 
target average firing rates, with $N=1000,K=100,f_{exc} = 0.5$ and $\delta_0=10^{-9}$ 
in all cases. Presented curves are the average of $10^4$ random 
perturbations. Inset: The limit of dynamical predictability 
$T_p$ defined as a time needed for an error to reach 
$98\%$ of the saturation level.}
\label{fig_avPredLyapun}
\end{figure}

We have evaluated the nonlinear finite 
time Lyapunov exponent (FTLE), which measures 
the short-term growth rate of initial perturbations 
without linearization of the time evolution equations 
\citep*{Ding2007}.

In practice the FTLE is estimated by considering a 
small perturbation of the trajectory 
$\vec{z}(t) \in \mathbb{R}^{3N}$ along a randomly selected 
direction $\vec\delta(0) \in \mathbb{R}^{3N}$, 
and following the deviation of the perturbated
trajectory from the reference orbit, that is 
$\vec\delta(\tau) = \vec{z}^\prime(t+\tau)-\vec{z}(t+\tau)$. 
The FTLE is the obtained as 
$\lambda(\vec{z}(t),\delta_0,\tau) = 
\frac{1}{\tau}\ln\frac{\delta(\tau)}{\delta_0}$, where 
$\delta(\tau) = ||\vec{\delta}(\tau)||$ and $\delta_0 = ||\vec\delta(0)||\ll 1$.
The FTLE depends on the starting point $\vec{z}(t)$ 
of the initial perturbation and on the size  
of the initial displacement $\delta_0$. 
The mean FTLE $\bar{\lambda}(\delta_0,\tau)$, 
which is independent from the starting point  
$\vec{z}(t)$ is evaluated by taking the average of the FTLEs 
over various points along the trajectory, 
$$\bar{\lambda}(\delta_0,\tau) = 
\langle \lambda(\vec{z}(t),\delta_0,\tau) \rangle_t = 
\frac{1}{n}\sum_{t=1}^n\lambda(\vec{z}(t),\delta_0,\tau)~.
$$
The mean FTLE still depends on the initial displacement 
$\delta_0$. If $\delta_0$ is chosen to be very small one 
observes initially an exponential growth of the perturbation. 
For this time period the mean FTLE is essentially constant 
and reduces to the largest Lyapunov exponent. The growth 
of the perturbation eventually enters a nonlinear phase, 
which is maintained until the deviation from the reference 
orbit reaches a saturation value. Note that FTLE 
($\lambda(\vec{z}(t),\delta_0,\tau)$) is not necessarily 
positive for all $\tau$ and $\vec{z}(t)$, implying
that an initial deviation can converge back towards 
the reference trajectory.   

When analysing the changes of the dynamical behavior 
as we reduced $\mu$, while keeping $f_{exc}$ 
constant at $0.5$, we found that when $\mu$ is 
in the range which corresponds to small values of 
$\langle D_{\lambda} \rangle$ 
(see lower right part of Fig.~\ref{fig_meanKLlinks}), 
the dynamical behavior is in a pure chaotic dynamical 
state with the constant part of the mean FTLE 
$\bar{\lambda}(t)$ in the range $[0.05, 0.1]$~. In this 
phase the FTLE is positive for all $\tau$ and $\vec{z}(t)$, 
thus small initial displacements diverge along every 
point of the orbit $\vec{z}(t)$.  

As the target mean firing rate is decreased down to $\mu = 0.2$ 
we observe a kink in the FTLE, with a transition to
a second linear time development, see Fig.~\ref{fig_avPredLyapun}. 
The manifestation of the kink corresponds to the occurrence 
of periods of quasi-constant activity as seen in the bottom 
plot of Fig.~\ref{fig_acti}. This laminar periods are 
characterised by a negative local Lyapunov exponent, 
that is small perturbations are suppressed during the laminar 
periods, leading however to a growth of the perturbation 
during the periods of chaotic bursting.

In this intermittent bursting phase the short term chaotic 
behavior ($t \lesssim 200$) describes the repulsion of two 
initially close trajectories mainly during the bursting regime. 
The long term behavior ($t \gg 200$) is also chaotic, as a 
consequence of the intermittent chaotic bursts. The change in the 
growth of perturbations (see Fig.~\ref{fig_avPredLyapun}) results
from the interplay of the distinct characteristic timescales of 
the intrinsic variables ($a_i$ and $b_i$) and the firing rates
($y_i$), with the first being slow and the 
later being fast variables \citep*{Boffetta1998}.    
The occurrence of transiently stable periods of activity
also leads to an increase in the time $T_p$, measuring 
the limit of dynamic predictability, as shown in the inset of 
Fig.~\ref{fig_avPredLyapun}. The duration of 
predictability $T_p$ is defined here as the time needed 
for a perturbation to reach $98\%$ of the saturation 
level \citep{Ding2007}. 
%

\begin{figure}[t]
\centerline{
\includegraphics[width=1\textwidth]{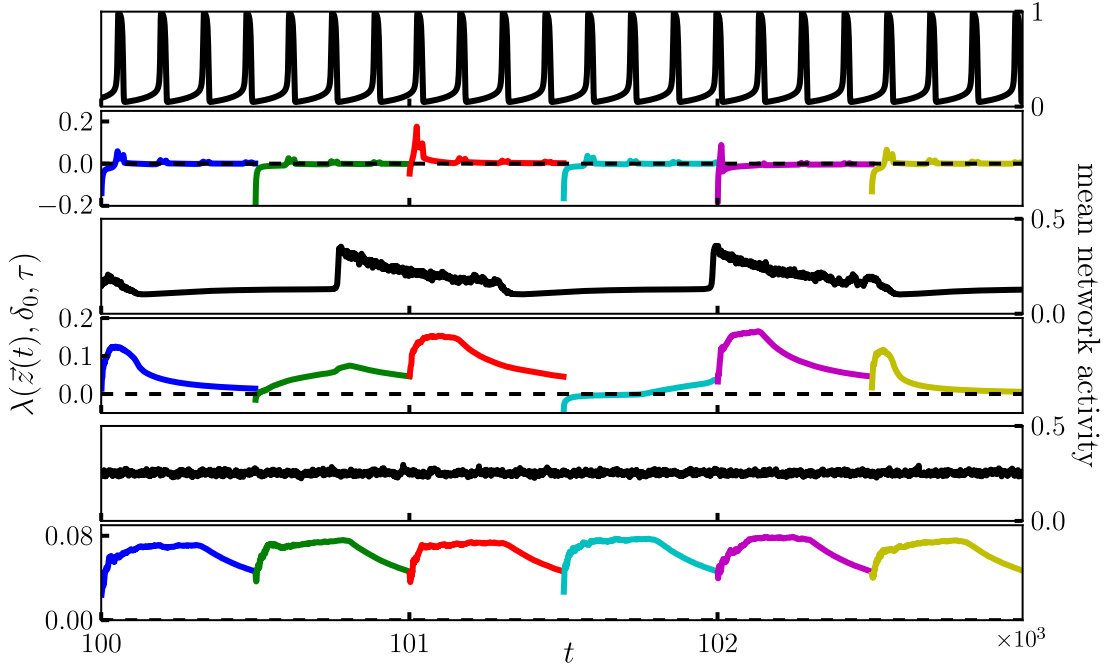}
}
\caption{Position dependent finite time Lyapunov 
exponent $\bar{\lambda}(\vec{z}(t),\delta_0, \tau)$ (see section \ref{sec_ftle}) 
along the orbit. We considered three cases: (bottom) a chaotic 
phase with $\mu=0.25$ and $f_{exc}=0.5$, (middle) intermittent-bursting 
phase with $\mu=0.15$ and $f_{exc}=0.5$, (top) synchronised oscillatory 
phase with $\mu = 0.15$ and $f_{exc}=0.7$. The initial displacement 
$\delta_0$, number of neurons $N$ and connectivity $K$ were set 
to $10^{-9}, 10^3$ and $10^2$, respectively.  
}
\label{fig_ftle}
\end{figure}

The dependency of the FTLE on the position of a perturbation 
along the orbit in phase space is presented in 
Fig.~\ref{fig_ftle}. We compared the FTLE, that is 
$\lambda(\vec{z}(t),\delta_0,\tau)$, as estimated from six 
different points along the trajectory $\vec{z}(t)$,
for all three dynamical regimes. The FTLE is then
evaluated for 500 consecutive timesteps and after the
$500\mbox{th}$ timestep a new perturbation is
introduced. In the pure chaotic dynamical regime 
the FTLE is positive for all given initial perturbation 
points.  While, in intermittent bursting regime one can 
also notice negative values of the FTLE when the perturbation 
is initiated within the laminar period. In contrast, 
perturbations starting during the periods of bursts 
leads to a strictly positive FTLEs. In the oscillatory 
regime the FTLE is negative along the orbit and, similar 
to the single neuron case, the trajectory is unstable during 
the fast transitions from low to high activity states 
(and vice versa), which results in sharp, positive valued, 
spikes in the FTLE.

Chaotic dynamics is also observed in the non-adapting limit with
$\epsilon \to 0$, whenever the static values of $a_i$ 
are above the critical value. This is in agreement with 
the results of a large-$N$ mean field analysis of an 
analogous continuous time Hopfield network~\citep*{Sompolinsky1988a}.
Subcritical static $a_i$ lead, on the other hand, 
to regular dynamics controlled by point attractors.

\section{Discussion}\label{sec_discussion}

Our results show that the introduction of intrinsic plasticity 
in random recurrent neural networks results in ongoing 
and self-sustained neural activities with non trivial 
dynamical states. For large networks we have observed, 
depending on the specified parameters, three
self-organized distinct phases. The network parameters
include the fraction of excitatory connections, 
the average connectivity
and the target average firing rate. These results show that
non-synaptic adaptation plays an important role in the formation
of complex patterns of neural activity.

An important part of the sensory signals an organism receives
result from the reactions of the environment to the
motor actions taken by the organism itself. 
The complexity of this portion of sensory inputs will then
depend on the complexity of the organism's own behavior.
This consideration indicates that self-generated and
autonomously sustained neural activity is important for
the generation of non-trivial behavioral patterns.
It is hence more likely that an animal will start an
explorative behavior if the brain, supporting the body, 
is able to maintain a change in sensory input. In other words, 
if the dynamics of a neural controller would approach, in the
absence of sensory inputs, a state of stable and constant 
activity, it would be capable of generating 
only trivial motor actions.

As mentioned in the introduction, synaptic plasticity 
alone drives the dynamics of a recurrent network generically
toward a frozen state, independent of the presence or absence
of sensory input \citep{Siri2007a}. Synaptic plasticity
is thus in general, for non-spiking neurons, not sufficient 
for achieving self sustained activity, a likely 
essential precondition to complex behavioral patterns.

The relevance of critical brain dynamics for both
non-linear sensory processing \citep*{kinouchi2006}
and for self-sustained neural computation is being
investigated intensively. \citet*{Levina2007,Levina2009} have
demonstrated that critical neural activity can be achieved when the
depletion of synaptic vesicles is included into the dynamics
of membrane potential. Under this setup, they observe power
law scaling of avalanches formed by the activity of spiking neurons,
a result in agreement with experimental observations 
\citep[e.g.,][]{Chialvo2010},
which are supportive of the notion that the brain works in critical
regime. However, without any form of adaptation to varying 
sensory stimuli, neurons would perform only trivial computations 
and criticality, or other complex activity patterns, 
would generically not arise. 
Thus, to properly understand the brain dynamics and cognitive 
processes, one must include various forms of 
plasticity \citep{Triesch2005,Triesch2007}.

Here we showed that intrinsic or non-synaptic plasticity
will drive a system of recurrently interacting neurons,
under certain quite general conditions 
(network connectivity, ratio of excitation versus inhibition), 
towards a chaotic phase. Synaptic adaption rules, on
the other side, are known to generically drive recurrent 
neural networks into a subcritical or frozen
state. Our results hence indicate that self-organization 
of neural network dynamics into a critical regime
could occur whenever intrinsic and synaptic plasticity 
are both present and relevant. Critical neural dynamics
would then result from the interplay between synaptic
and non synaptic adaption processes.

An analogous line of arguments has been brought
forward by \citet*{Der2006}, by
demonstrating the relevance of self-organized criticality 
for the emergence of exploratory behavior in 
autonomous agents. Optimal predictability of the sensory-motor 
cycle is achieved when the neural controller works 
in a critical regime \citep{Der2006}. We believe that 
self-organized criticality in biologically inspired autonomous 
recurrent neural networks will exhibit similar patterns of complex 
behavior. Certainly, the complex behavior should persist when 
including the interaction between an agent and the environment, 
as we plan to do in future investigation.
%

\bibliographystyle{natbib}

\end{document}